\documentclass[12pt]{article}
\usepackage{epsfig}
\usepackage{citesort}

\newcommand{\eq}{\begin{equation}}
\newcommand{\eqx}{\end{equation}}
\newcommand{\eqn}{\begin{eqnarray}}
\newcommand{\eqnx}{\end{eqnarray}}

\newcommand{\dt}{\delta}
\newcommand{\Th}{\theta}
\newcommand{\om}{\omega}
\newcommand{\w}{\om^{\prime}}
\newcommand{\ww}{\om^{\prime 2}}
\newcommand{\www}{\om^{\prime {2 \over 3} }}

\newcommand{\nin}{\noindent}
\newcommand{\lam}{\lambda}
\newcommand{\pau}{$\,\,$}
\newcommand{\seld}{\frac{d\sigma_{el}}{d\Th}(E)}
\newcommand{\tel}{\sigma_{el}(E)}
\newcommand{\eps}{\epsilon}

\begin{document}
\nin
\begin{center}
{\Large \bf Auger-electron cascades in diamond and
amorphous carbon}\\
\vspace{10mm}

{\Large Beata~Ziaja $^{\ast,\,\dag,\,\ddag}$}
{\Large David~van der Spoel $^{\ast}$} ,
{\Large Abraham~Sz\"{o}ke $^{\ast,\,\S}$} ,
{\Large Janos~Hajdu $^{\ast}$} \footnote{e-mail:ziaja@tsl.uu.se,
~spoel@xray.bmc.uu.se,~szoke1@llnl.gov,~hajdu@xray.bmc.uu.se}

\vspace{3mm}
           $^{\ast}$  \it Department of Biochemistry, Biomedical Centre,
                   \it Box 576, Uppsala University, S-75123 Uppsala, Sweden\\
\vspace{3mm}

           $^{\dag}$ \it Department of Theoretical Physics,
	               Institute of Nuclear Physics,
                      \it Radzikowskiego 152, 31-342 Cracow, Poland\\
\vspace{3mm}
           $^{\ddag}$ \it High Energy Physics, Uppsala University,
	               P.O. Box 535, S-75121 Uppsala, Sweden
\vspace{3mm}

           $^{\S}$ \it Lawrence Livermore National Laboratory, Livermore,
	          CA 94551, USA\\
\end{center}

\vspace{5mm}
\nin
{\bf Corresponding author:}\\
Janos Hajdu, Department of Biochemistry, Biomedical Centre,\\
Box 576, Uppsala University, S-75123 Uppsala, Sweden\\
Tel:+4618 4714999, Fax:+4618 511755, E-mail:hajdu@xray.bmc.uu.se \\ \\
\nin
{\bf Manuscript information:}\\
Number of pages:18, Number of figures:9, Number of tables:0\\ \\
%\nin
%{\bf Word and character counts:}\\
%Abstract:1055, Total number of characters in the paper: about 34134\\ \\
%\nin
%{\bf Abbreviations footnote:}\\
%DIMFP-differential inverse mean free path,\\
%ELF-energy loss function,\\
%EMFP-elastic mean free path,\\
%FEL-free-electron laser,\\
%IMFP-inelastic mean free path,\\
%MC-Monte-Carlo,\\
%MFP-mean free path,\\
%QM-quantum-mechanical,\\
%TPP-2$\,\,$-Tanuma, Powell and Penn model.
%\\ \\
\nin
{\bf Abstract:}
We have analyzed the cascade of secondary electrons in diamond and
amorphous carbon generated by the thermalisation of a single Auger
electron. The elastic electron mean free path was calculated as a function
of impact energy in the muffin-tin potential approximation. The
inelastic scattering cross section and the energy loss of the electron
(expressed in terms of differential inverse mean free path) were
estimated from two "optical" models, that utilise the measured dielectric
constants of the materials. Using these data, a Monte-Carlo model
describing the time evolution of the cascade was constructed. The results
show that at most around $20-40$ secondary cascade electrons are
released by a single Auger electron in a macroscopic sample of diamond or
amorphous carbon. Consideration of the real band structure of diamond reduces
this number further. 
The release of the cascade electrons happens within the first
$100$ femtoseconds after the emission of the primary Auger electron. The
results have implications to planned experiments with femtosecond X-ray
sources.
\vfill
\clearpage
\setcounter{page}{1}
%%%%%%%%%%%%%%%%%%%%%%%%%%%%%%%%%%%%%%%%%%%%%%%%%%%%%%%%%%%%%%%%%%%%%%%%%%%%%
%%%%%%%%%%%%%%%%%%%%%%%%%%%%%%%%%%%%%%%%%%%%%%%%%%%%%%%%%%%%%%%%%%%%%%%%%%%%%
\section{Introduction}

%%%%%%%%%%%%%%%%%%%
Radiation damage prevents structure determination of single biomolecules
and other non-repetitive structures at high resolutions in standard
electron or X-ray scattering experiments \cite{Henderson95}. Cooling can
slow down sample deterioration, but it cannot eliminate damage-induced
sample movement within the time needed to complete conventional
measurements \cite{Henderson95,Henderson90,Nave95}. 
Emerging new X-ray sources, like free-electron lasers (FEL) 
\cite{Hajduslac,Hajdutesla}, will offer new possibilities in imaging. Analysis 
of the dynamics of damage formation on a sample in an X-ray FEL beam 
suggests that the conventional damage barrier
(about 200 X-ray photons/{\AA}$^2$ at 12~keV energy) \cite{Henderson90},
may be extended substantially at very high dose rates and very short
exposure times \cite{l1,Hajdu2000}. A new dynamic barrier of
radiation tolerance has been identified at extreme dose rates and ultra
short exposure times \cite{Hajduslac,Hajdutesla,l1,Hajdu2000}. 
This barrier is several orders of magnitude
higher than previous theoretical limits in conventional experiments. The
calculations show that at
these extremes, sections of molecular transforms from single
macromolecules may be recorded without the need to amplify scattered
radiation through Bragg reflections \cite{Hajduslac,l1}. 

At $1$ \AA $\,$ wavelength, about nine-tenth of the interacting photons will deposit
energy into a biological sample, causing damage mainly through the
photoelectric effect. The departing photoelectron leaves a hole in a low
lying orbital, and an upper shell electron falls into it. This electron
may either emit an X-ray photon to produce X-ray fluorescence or may give 
up its energy to another electron, which is then ejected from the ion as 
an Auger electron. The probability of fluorescence
emission or Auger emission depends on the binding energy of the
electron. In biologically relevant light elements, the predominant
relaxation process ($>99$\%) is through Auger emission, and most
photoelectric events ultimately remove two electrons from these elements
(C, N, O, S). The two electrons have different energies and leave
the atom at different times (for a more detailed description, see \cite{l1,l3,l5}). 

In very small samples (like atoms and single molecules), the primary
photoelectrons and the Auger electrons may escape from the sample without
further interactions. However, in larger samples, these electrons will
become trapped and thermalised. Thermalisation involves inelastic
electron-atom interactions, producing secondary cascade electrons. 
Here we analyze the specific contribution of Auger electrons to the
ionization of a macroscopic sample through secondary cascade processes. 
We selected two different carbon compounds (diamond and amorphous carbon) 
as models for the calculations. 

Auger electrons and photoelectrons propagate through the medium
in a different manner.
Their de Broglie wavelengths are $\lam_{Auger}\approx0.8$ \AA \pau 
and $\lam_{photoel}\approx0.1$ \AA, respectively, and 
$\lam_{Auger}$ is comparable with atomic size.
This implies that Auger electrons interact multiply with
neighbouring atoms, while moving through the system of atoms
in the solid \cite{l6}. Moreover, since the energy of Auger electrons is low
(around 0.25~keV), the interaction potential must include a non-local exchange 
term which makes accurate description of the interaction complicated. 
In contrast, photoelectrons propagate almost freely through the medium, 
and their interaction with (single) atoms in the medium is well described 
by the Born approximation \cite{l6prim,l6bis}.
Therefore, in samples of intermediate size the low energy
Auger electrons are more likely to cause significant ionization than the 
higher energy photoelectrons. The energy dependence of the mean free
path (MFP) of electrons \cite{l7,l14prim} in carbon implies that the MFP 
of a photoelectron is of the order of hundred \AA ngstroms whereas the MFP
for the  Auger electron is only a few \AA ngstroms ($\geq 4$ \AA).
%(cf.\ Fig. \ref{f1}, \ref{f2}, \ref{f3}). 
This implies that in samples of
intermediate size a photoelectron scatters only a few times before leaving
the interaction region, while the Auger electron will have multiple 
interactions.

In section {\bf 2} we quantify the elastic and inelastic interactions of 
Auger electrons with atoms within a solid. Using a Monte-Carlo (MC) simulation 
we then model the secondary electron cascade caused by inelastic interactions 
of the primary electron and subsequent secondary electrons with atoms. 
In section {\bf 3}, the results of 500 computer simulations of different
cascades are presented. These results give the estimated average ionization rate
as a function of time.
Finally, in section {\bf 4} we list our conclusions. 

%%%%%%%%%%%%%%%%%%%%%%%%%%%%%%%%%%%%%%%%%%%%%%%%%%%%%%%%%%%%%%%%%%%%%%%%%%
%FIGURE 1
\begin{figure}[t]
      \centerline{\epsfig{figure=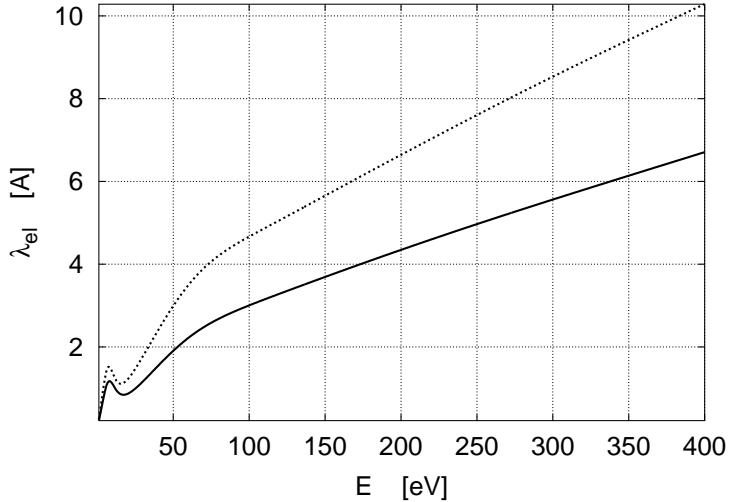,width=10cm}}
      \caption{{\footnotesize Elastic mean free path ($\lambda_{el}$ or EMFP) 
of electrons in diamond or amorphous carbon plotted as a function of electron 
energy $E$. Solid line corresponds to the EMFP of electron in diamond, dotted 
line shows the EMFP of electron in amorphous carbon.  
}}
\label{f1}
\end{figure}
%FIGURE 4
\begin{figure}[t]
      \centerline{\epsfig{figure=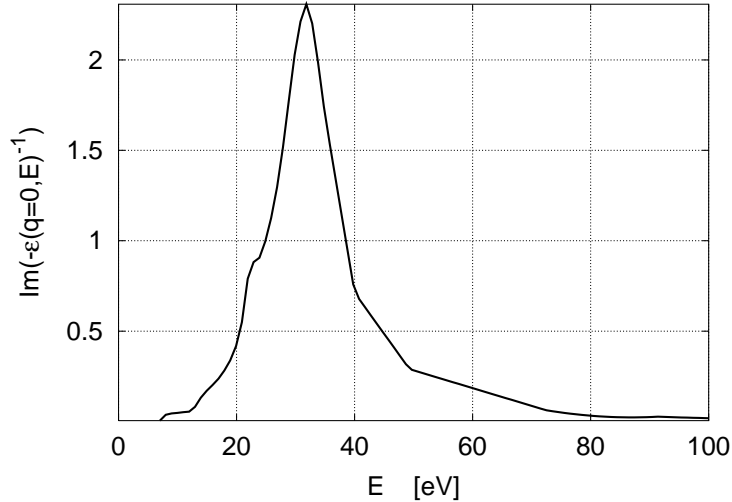,width=10cm}}
      \caption{{\footnotesize Energy loss function, $Im[-\eps(q=0,E)^{-1}]$
for diamond plotted as a function of photon impact energy $E$.
}}
\label{f4}
\end{figure}
%
%FIGURE 5
\begin{figure}[t]
      \centerline{\epsfig{figure=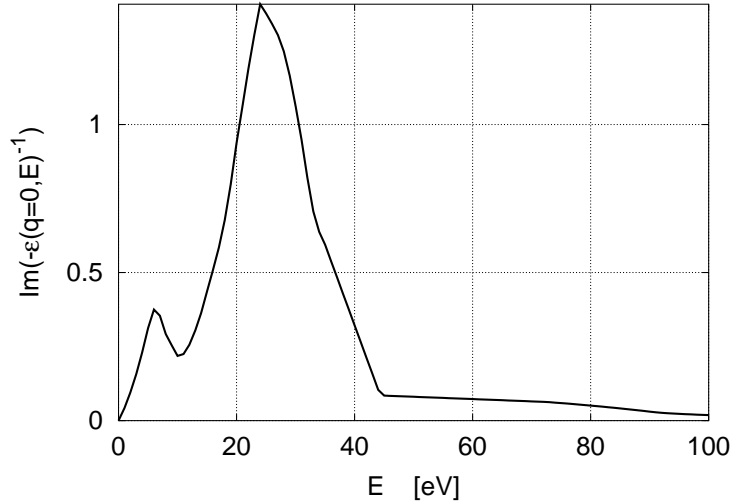,width=10cm}}
      \caption{{\footnotesize Energy loss function, $Im[-\eps(q=0,E)^{-1}]$
for amorphous carbon plotted as a function of photon impact energy $E$.
}}
\label{f5}
\end{figure}
%
%FIGURE 2
\begin{figure}[t]
      \centerline{\epsfig{figure=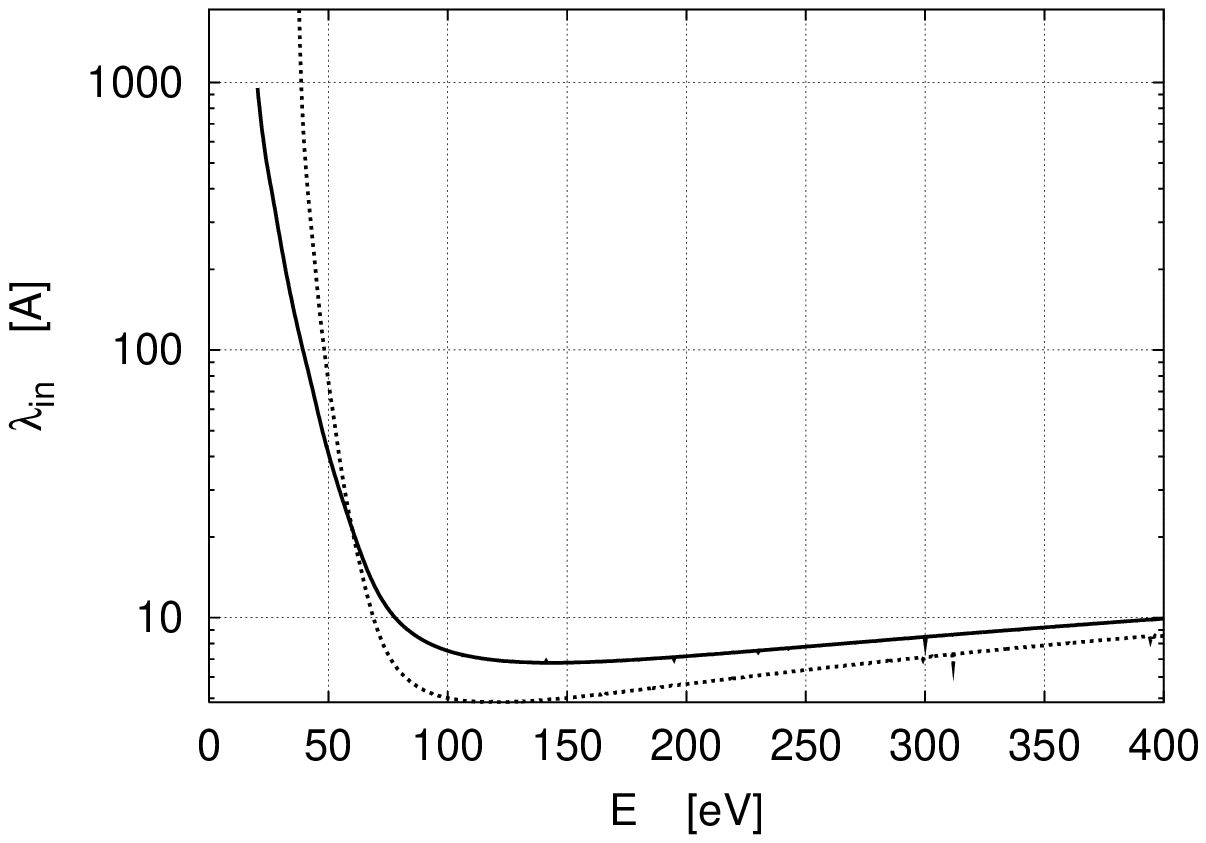,width=10cm}}
      \caption{{\footnotesize Inelastic mean free path ($\lambda_{in}$ or IMFP) 
of electrons in diamond plotted as a function of electron energy $E$.
Solid line corresponds to the IMFP calculated from Ashley's model
(\ref{ashleyfin}), dotted line shows the IMFP calculated from the TPP-2 model 
(\ref{tanumaf}).
}}
\label{f2}
\end{figure}
%FIGURE 3
\begin{figure}[t]
      \centerline{\epsfig{figure=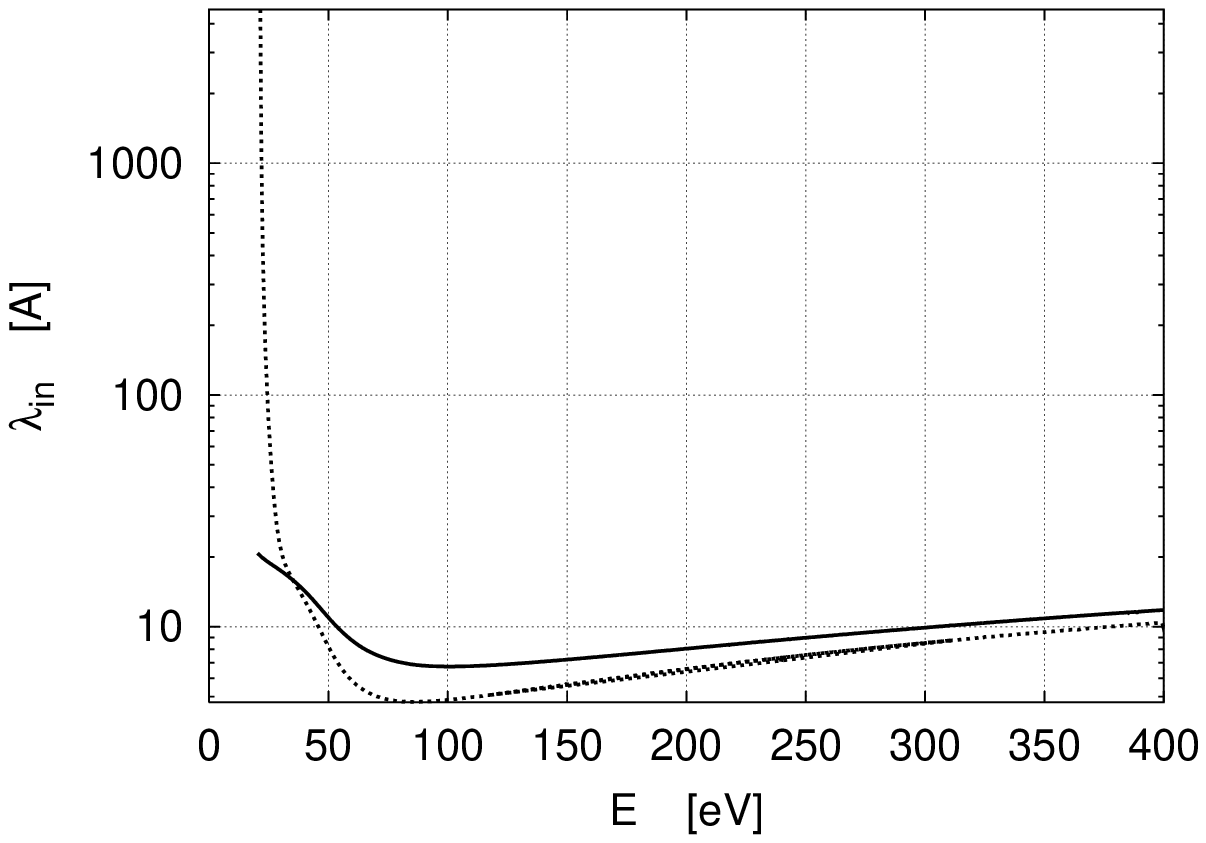,width=10cm}}
      \caption{{\footnotesize Inelastic mean free path ($\lambda_{in}$ or IMFP) 
of electrons in amorphous carbon plotted as a function of electron energy $E$.
Solid line corresponds to the IMFP calculated from Ashley's model
(\ref{ashleyfin}), dotted line shows the IMFP calculated from the TPP-2 model 
(\ref{tanumaf}).
}}
\label{f3}
\end{figure}
%
%FIGURE 6
\begin{figure}[t]
      \centerline{\epsfig{figure=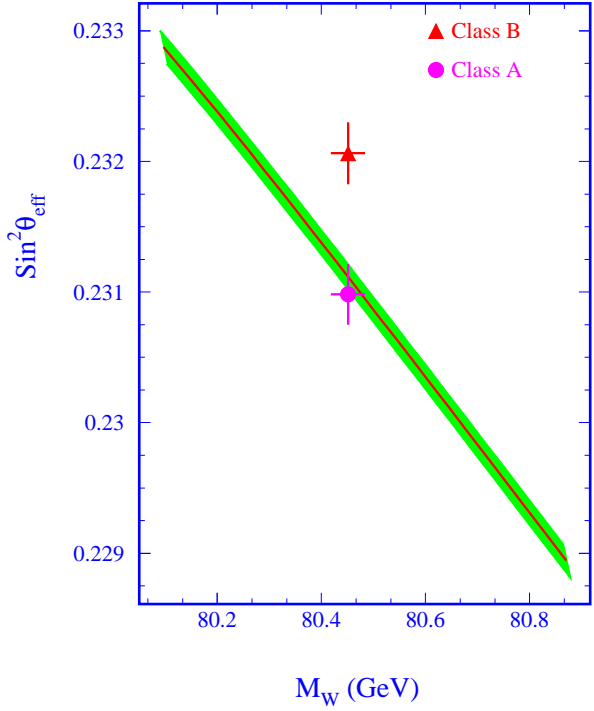,width=10cm}}
      \caption{{\footnotesize Example of an electron path in a solid.
}}
\label{f6}
\end{figure}
%
% FIGURE 7
\begin{figure}[t]
\epsfig{width=13cm, file=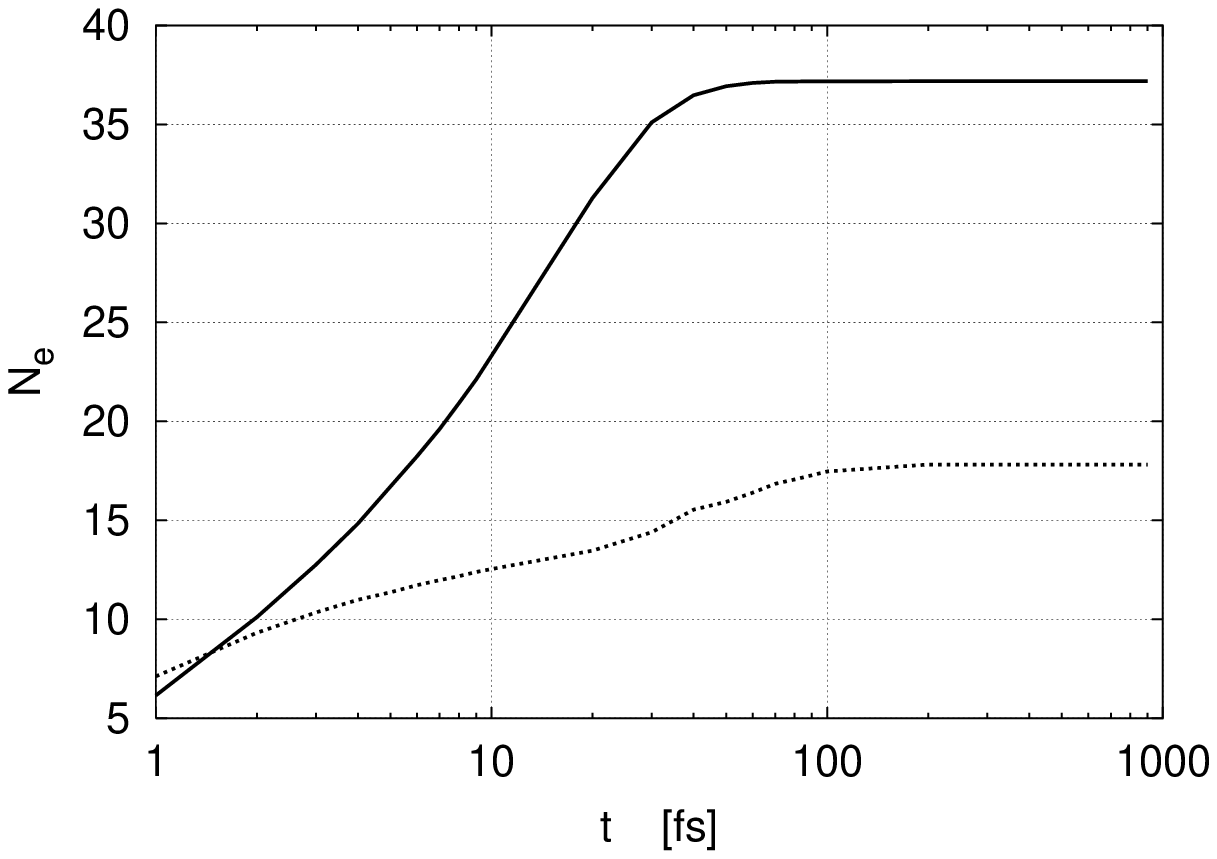}\hfill\mbox{} \\
\epsfig{width=13cm, file=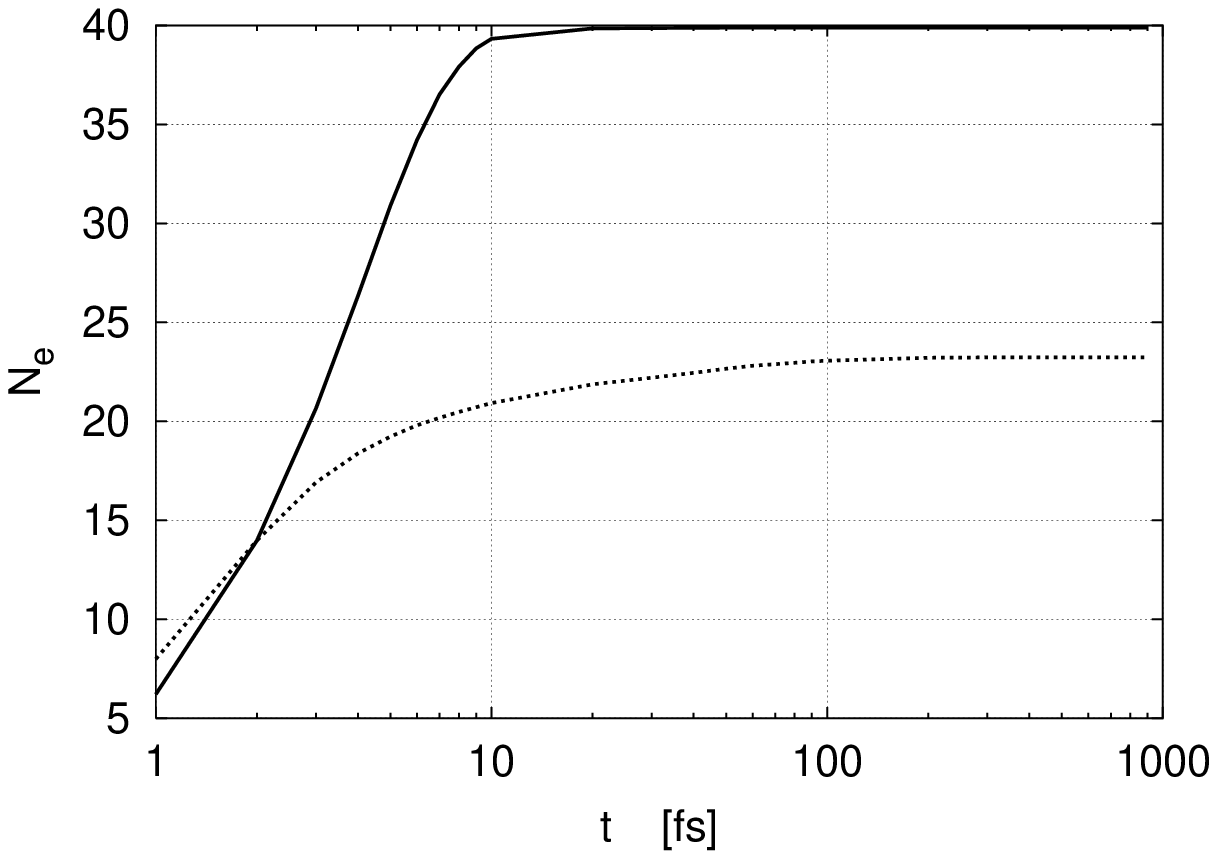}\hfill\mbox{} \\
\caption{The average ionization rate, $N_{e}$, plotted as a function of time,
$t$, for diamond (upper plot) and for amorphous carbon (lower plot). 
The energy of the primary Auger electron was $E=250+E_F$ eV, where $E_F$
is the Fermi energy~: $E_{F}\approx 29$ eV for diamond and $E_{F}\approx 21$ eV 
for amorphous carbon. Solid lines correspond to the average ionization estimated 
from Ashley's model, dotted lines show the average ionization calculated from
the TPP-2 model. 
}
\label{f7}
\end{figure}
%
% FIGURE 8
\begin{figure}[t]
\epsfig{width=13cm, file=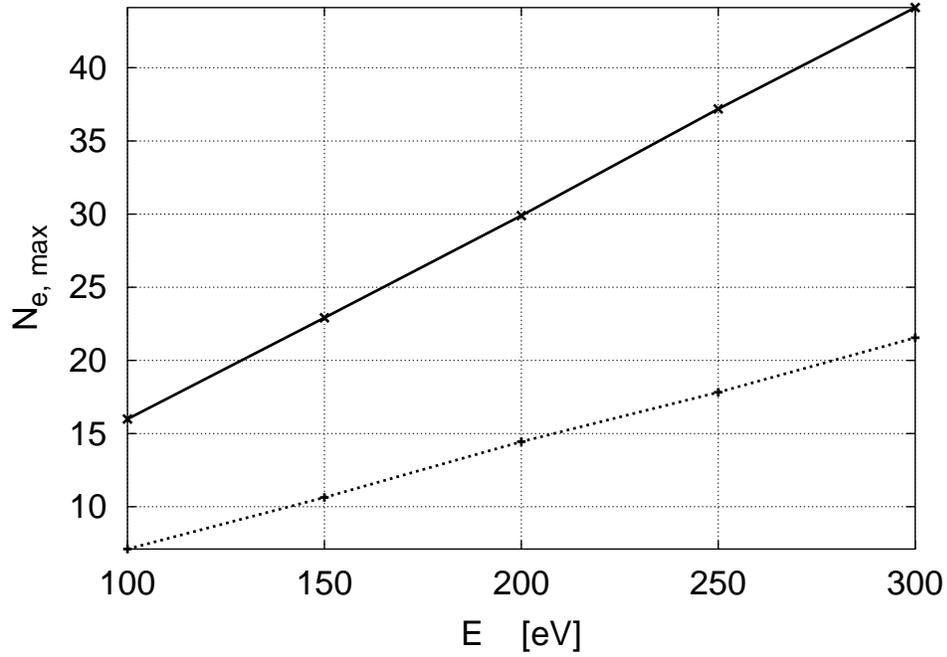}\hfill\mbox{}\\
\epsfig{width=13cm, file=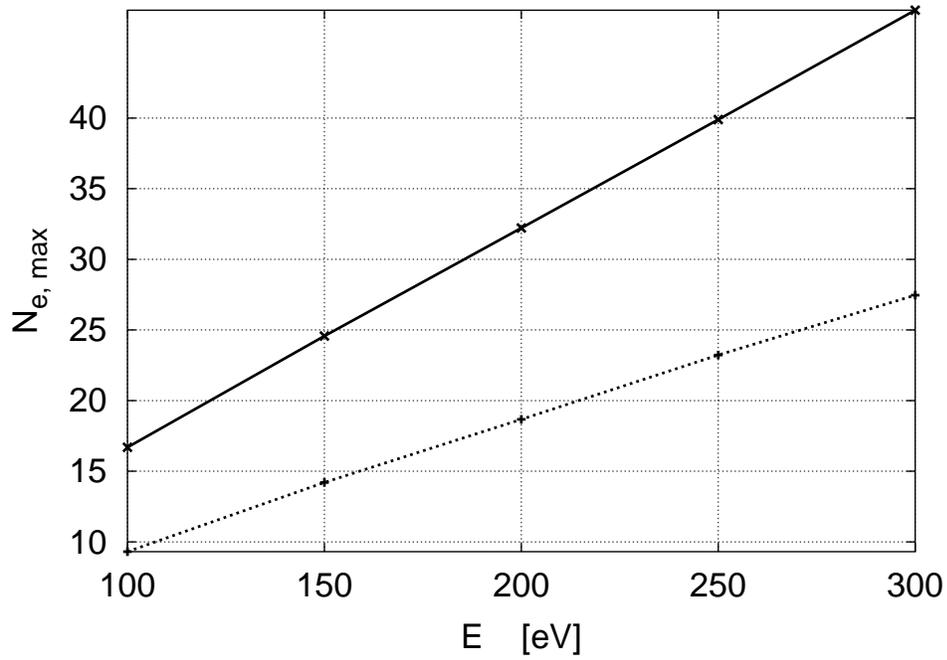}\hfill\mbox{}\\
\caption{Maximal ionization rate, $N_{e,max}$, plotted as a 
function of the energy of the primary Auger electron $E=E^{\prime}+E_F$ 
($E^{\prime}=100,150,200,250,300$ eV) for diamond (upper plot) and for amorphous 
carbon (lower plot). The corresponding Fermi energies are~: $E_{F}\approx 29$ 
eV for diamond and $E_{F}\approx 21$ eV for amorphous carbon. Solid lines 
show the maximal ionization estimated from Ashley's model, dotted 
lines show maximal ionization calculated from the TPP-2 model.
}
\label{f8}
\end{figure}
%
% FIGURE 9
\begin{figure}[t]
\centerline{\epsfig{width=13cm, file=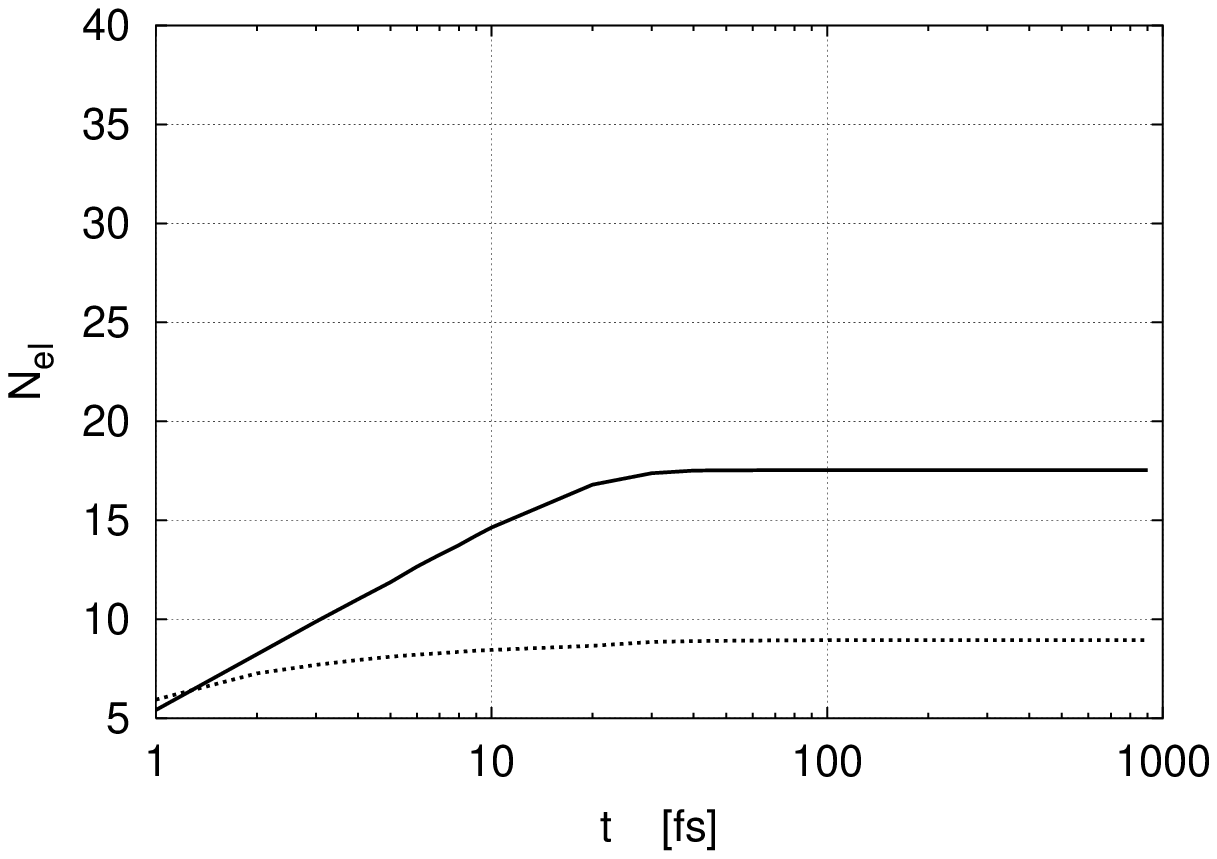}}
\caption{The average ionization rate, $N_{e}$, plotted as a function of time,
$t$, for diamond after the real band structure of diamond at $T=300$ K was 
included into the model. The energy of the primary Auger electron was 
$E=250+E_F$ eV, where $E_F$ is the Fermi energy~: $E_{F}\approx 29$ eV for 
diamond. The energy gap at $T=300$ K equals $E_{gap}=5.46$ eV.
Solid lines correspond to the average ionization estimated 
from Ashley's model, dotted lines show the average ionization calculated from
the TPP-2 model. 
}
\label{f9}
\end{figure}
%
%%%%%%%%%%%%%%%%%%%%%%%%%%%%%%%%%%%%%%%%

%%%%%%%%%%%%%%%%%%%%%%%%%%%%%%%%%%%%%%%%%%%%%%%%%%%%%%%%%%%
\section{Secondary electron cascade in a solid}
%%%%%%%%%%%%%%%%%%%%%%%%%%%%%%%%%%%%%%%%%%%%%%%%%%%%%%%%%%%

Our study of the secondary electron effects was performed for two forms
of carbon: diamond ($\rho=3.51\,g/cm^3$) and graphite-like amorphous
carbon ($\rho=2.21\,g/cm^3$). 

Low energy electrons ($E\approx250$ eV) may undergo elastic and inelastic 
collisions with atoms (electrons and nuclei) in a solid. Since the corresponding 
electron wavelength is comparable with atomic dimensions, multiple scattering of
the electron \cite{l8} on neighbouring atoms have to be calculated 
quantum-mechanically (QM). The QM exchange terms must then be incorporated
into the interaction potential.

%%%%%%%%%%%%%%%%%%%%%%%%%%%%%%%%%%%%%%%%%%%%%%%%%%%%%%%%%%%
\subsection{Elastic scattering}
%%%%%%%%%%%%%%%%%%%%%%%%%%%%%%%%%%%%%%%%%%%%%%%%%%%%%%%%%%%

Calculation of elastic scattering amplitudes and angular
distributions can be done accurately by the partial wave expansion technique
\cite{l6prim,l6bis}. In particular, the differential elastic cross-section
$\displaystyle \seld\,\,$ for scattering of an electron
on the atom is expressed, using the phase shift $\displaystyle \dt_l$ of each partial wave
as follows \cite{l6prim,l6bis}~:
\eq
\seld=\frac{2\pi}{k^2}\mid
\sum_{l=0}^{\infty}\,(2l+1)\sin(\dt_l)\,P_l(\cos(\Th))
\mid^2,
\label{seld}
\eqx
where $k$ is the wave number, corresponding to the electron impact energy $E$,
$P_l(\cos(\Th))$ denotes the Legendre polynomial of order $l$, $\Th$ is the
scattering angle and the sum goes over all partial wave contributions
$l=0,\ldots\infty$. The total elastic cross-section 
$\displaystyle \tel$ may be obtained from
(\ref{seld}) by integration over $\Th$. The corresponding elastic
mean free path, (EMFP) $\lambda_{el}$ \cite{l7,l15} can be calculated as~:
\eq
\lambda_{el}^{-1}(E)= N\,\sigma_{el}(E),
\label{lel}
\eqx
where $N$ denotes the atomic density in the solid. In order to obtain the 
phase shifts, $\dt_l$ in (\ref{seld}), one should solve the respective radial wave
equations for each partial wave with the approximate form of the exchange
potential.
To perform these calculations we used programs from the Barbieri/Van Hove
Phase Shift package \cite{l12}. First we determined the radial charge density 
for a free atom, then calculated the radial muffin-tin potential \cite{l6,l8} 
for atoms embedded in a solid (using various approximations to the exchange 
potential), and finally derived phase shifts from the muffin-tin potential. 
Multiple elastic scattering within a finite cluster, provided the resulting 
amplitude was large enough, was included in the calculations. Figure \ref{f1}
shows the resulting EMFP for diamond and amorphous carbon.
For large energies, the EMFPs for both diamond and amorphous carbon increase
linearly with electron impact energy. They also scale properly with the medium
density, $\displaystyle \frac{\lambda_{el,diamond}(E)}{\lambda_{el,carbon}(E)}\approx
\frac{\rho_{carbon}}{\rho_{diamond}}$ (cf.\ \cite{l14,l14prim}).
With decreasing energy the EMFPs decrease monotonically until they show 
oscillatory features due to interference between low-order
scattering waves.

%%%%%%%%%%%%%%%%%%%%%%%%%%%%%%%%%%%%%%%%%%%%%%%%%%%%%%%%%%%
\subsection{Inelastic scattering}
%%%%%%%%%%%%%%%%%%%%%%%%%%%%%%%%%%%%%%%%%%%%%%%%%%%%%%%%%%%

An accurate treatment of inelastic atom-electron collisions in a solid is more
difficult, especially in the case of low energy Auger electrons when multiple
scattering is important. In fact, a fully rigorous method for
including inelastic scattering  is not available so far.
Following  Fermi's work \cite{l16}, the passage of a fast charged particle was
treated through the linear perturbation caused by its electric field in
the solid.
Subsequent developments \cite{l17,l17prim,l17bis,l17tert,l17quar,l18} made it
possible to extend the dielectric formulation in order to provide a more
comprehensive description of quantum-mechanical effects in solids.

Generally speaking, the linear response of a solid is described by a generalized
dielectric constant, $\eps({\bf q},\om)$,
that depends both on momentum $\hbar{\bf q}$ and frequency $\om$.  In
quantum mechanics
$ \hbar \om$ corresponds to the energy transfer of the incident
charged particle to the
solid and $ \hbar {\bf q}$ to its momentum transfer.

It was shown \cite{l17tert} that the imaginary part of the dielectric constant,
$Im[-\eps({\bf q},\om)^{-1}]$,
determines the energy loss of the test charge per unit time,
${dE}/{dt}$, by the
formula, ${dE}/{dt}\sim\int dq \int d\om\,Im[-\eps({\bf q},\om)^{-1}]$.
Therefore
$Im[-\eps({\bf q},\om)^{-1}]$ is often called the energy loss function (ELF).
It satisfies the oscillator-strength sum rule \cite{l14}, that 
relates the total
energy loss to an effective number of free electrons per atom, $Z_{eff}$~:
\eq
Z_{eff}=\frac{2}{\pi \hbar^2 \Omega_P^2}\,\int_{0}^{\infty}dE\,E\,
Im[-\eps({\bf q},E/\hbar)^{-1}],
\label{oscs}
\eqx
where $\Omega_P=\sqrt{ ({4\pi n_a e^2})/{m_e} }$, $n_a=N_A\rho/ A$ is the 
density of atoms, $N_A$ is Avogadro's number, $\rho$ is the density of the solid,
$A$ is the atomic weight, and $E$ is the energy loss of the incoming test particle.

The energy loss of Auger electrons in a solid is dominated by the excitation of
plasmons.  At first, we expect this behavior in metals, where 
conduction electrons form a jellium-like plasma, but not in good insulators.
Nevertheless, in all solids the energy loss is dominated
by the excitation of valence electrons to the conduction band.  The excited
electron, in turn, interacts strongly with all other valence electrons.  The
resultant eigenstate is a plasma resonance. A more familiar result of similar
interactions among atoms in a solid is the formation of optical phonons. As
expected, the  plasmon interacts strongly with the incident Auger 
electron.  For a more quantitative explanation, let us examine the  
dielectric function $\eps({\bf q},\om)$.  It shows the importance of collective
modes for the energy loss of charged particles. If one rewrites, 
$\eps=\eps_1+i\eps_2$ then
$Im[-\eps^{-1}]={\eps_2}/({\eps_1^2+\eps_2^2})$. Since $\eps_2$ is small,
if $\eps_1$ goes to $0$ at a certain frequency $\om=\om_P$,
the ELF, $Im[-\eps^{-1}]$, peaks sharply at this frequency. This corresponds
to  excitation of plasma modes of frequency
$\om_P$ by the incoming particle.  Therefore, approximating the solid 
as a gas of free electrons, models the electron energy loss well. As the width
of the plasma resonance and its amplitude depend on the details of the plasmon
coupling and its decay, accurate results can be expected only from detailed 
simulations.

In this paper we apply the Lindhard dielectric function approach  together
with optical-data models. The approximation proved to work well
in free-electron-like materials where the ELF $Im[-\eps(0,\om)^{-1}]$
registered for incoming photons shows a dominant peak due to well-defined
volume plasmons \cite{l14,l19}.

Similarly as above, the response of the medium to a passing
electron of a given energy $\hbar\om$ and momentum $\hbar{\bf q}$ is then
described by a complex Lindhard dielectric function \cite{l17prim}
$\eps({\bf q},\om)$.
In general $\eps$ may be a tensor but it is assumed here that the medium is
homogeneous and isotropic. In this case, $\eps$ is a scalar function which depends only
on the magnitude of $\hbar{\bf q}$. The probability of an energy loss 
$\hbar\om$
per unit distance travelled by a non-relativistic electron of energy 
$E$,  i.e. the
differential inverse mean free path (DIMFP) $\tau(E,\om)$
\cite{l17prim,l17quar,l21,l21prim},
then reads~:
\eq
\tau(E,\om)=\frac{1}{\pi E a_0}\,\int_{q_-}^{q_+}\,
\frac{dq}{q}\,Im[-\eps(q,\om)^{-1}],
\label{tau}
\eqx
where $a_0$ is the Bohr radius, and~:
\eq
q_{\pm}=k\left(1\,\pm\,\sqrt{1-(\hbar\om/E)}\right)
\label{qpm}
\eqx
for $k$ denoting the wave number corresponding to electron impact energy $E$.
The expression for $q_{\pm}$ assumes that the energy and momentum transfer
for electron moving in the medium is the same as for a free particle in vacuum,
i.e. there is no effective mass assumed.
Integration of the DIMFP over the allowed values of $\om$ yields
the inelastic mean free path (IMFP) through~:
\eq
\lambda_{in}^{-1}(E)= \int\,d\om\,\tau(E,\om).
\label{lin}
\eqx

It follows from (\ref{tau}) that the only quantity needed to evaluate
$\tau(E,\om)$ and $\lambda_{in}(E)$ is the dielectric response function
$\eps(q,\om)$. However, most existing data on dielectric response functions 
were obtained from photon scattering on solids, for which  the momentum 
transfer is zero.
The problem is how to predict the dielectric response function with $q>0$, 
knowing only its optical limit ($q=0$) \cite{l21,l21prim}. For that purpose 
a phenomenological optical model approach was introduced, where
$Im[-\eps(q,\om)^{-1}]$ is expressed via the convolution of
$Im[-\eps(q=0,\om)^{-1}]$ with some profile function of $q$ and $\om$.

The two transparent optical models we apply hereafter were chosen to give
a reasonable estimate of ionization rate within the accuracy required for
our model. In what follows we will use  atomic units ($\hbar=e=m=1$)
if not stated explicitly.

The optical model by Ashley \cite{l21,l21prim} includes exchange 
between the incident electron and the electron in the medium modeled 
in analogy with the structure of the non-relativistic M\o{}ller cross-section~:
{\footnotesize
\eqn
\tau_A(E,\om)&=&{1 \over {2\pi E}}\, \int_0^{\infty}\, d\w \,\w \,
Im[-\eps(0,\om)^{-1}]\nonumber\\
&\times&\{F(E,\w,\om)+F(E,\w,E+\w-\om)-\sqrt{F(E,\w,\om)F(E,\w,E+\w-\om)}\},
\label{ashley}
\eqnx
}
where~:
{\footnotesize
\eq
F(E,\w,\om)={\bar \Theta}(\om-q_-^2/2-\w > 0)\,{\bar
\Theta}(\w+q_+^2/2-\om > 0)
{1 \over {\om(\om-\w)}},
\label{ashleyf}
\eqx
}
and ${\bar \Theta}$ is the step function. Substituting (\ref{ashleyf}) into
(\ref{ashley}) one obtains \cite{l21,l21prim}~:
{\footnotesize
\eqn
\tau_A(E,\om)&=&{1 \over {2\pi E}}\, \int_0^{\infty}\, d\w \,\w \,
Im[-\eps(0,\om)^{-1}]\nonumber\\
&\times&\left( {1 \over {\om(\om-\w)}}+{1 \over {(E+\w-\om)(E-\om)}}
-{1 \over \sqrt{\om(\om-\w)(E+\w-\om)(E-\om)}}\right)\nonumber\\
&\times&(\Theta_1(E,\w,\om)+\Theta_2(E,\w,\om)),
\label{ashleyfin}
\eqnx
}
where $\Theta_1$ and $\Theta_2$ restrict the integration region over
$\w$ and $\om$~:
{\footnotesize
\eqn
\Theta_1(E,\w,\om)&=& {\bar \Theta}(0\,<\,\om\,<\,E/2)\,\,
{\bar \Theta}\left(0\,<\,\w\,<\,2E (\om/E-1+\sqrt{1-\om/E})\right),\\
\Theta_2(E,\w,\om)&=& {\bar \Theta}(E/2< \om < 3E/4)\,\,
{\bar \Theta}\left(2\om-E\,<\,\w\,<\,2E (\om/E-1+\sqrt{1-\om/E})\right).
\label{ashleylim}
\eqnx
}
The Tanuma, Powell and Penn model (TPP-2) \cite{l14} was
adopted for calculating the DIMFP and IMFP of electrons
in a solid. We have not used the TPP-2 fit for IMFP calculation but derived
the DIMFP, and consequently IMFP, explicitly from statistical approximation
described in \cite{l14}. The DIMFP $\tau_T(E,\om)$ yields~:
\eqn
\tau_T(E,\om)&=&{1 \over {2\pi E}}\, \int_0^{\infty}\, d\w \,\w \,
Im[-\eps(0,\om)^{-1}]\nonumber\\
&\times&{1 \over
{\sqrt{c(\w)^2-\ww+\om^2}\left(\sqrt{c(\w)^2-\ww+\om^2}-c(\w)\right)} }
\nonumber\\
&\times&{\bar \Theta}(q_-^2/2 <\sqrt{c(\w)^2-\ww+\om^2}-c(\w)< q_+^2/2),
\label{tanumaf}
\eqnx
where $c(\w)=k_F(\w)^2/3$, and $k_F(\w)$ is the Fermi wave number for
the free-electron gas with plasma frequency equal to $\w$~:
\eq
k_F(\w)=\left(\frac{3\pi}{4}\right)^{\frac{1}{3}}\,\www.
\label{fe}
\eqx
The corresponding IMFP may be obtained after integrating (\ref{tanumaf}) over
$\om$, according to (\ref{lin}), taking into account the following
restrictions~:
\eqn
(q_-)^2/2 &<&\sqrt{c^2-\ww+\om^2}-c< (q_+)^2/2, \label{res1}\\
E-E_F   &<&\om,\label{res2}
\eqnx
where $E_F$ denotes the Fermi energy (see below). In particular,
restriction (\ref{res1}) implies $\w<\om$.
The energy loss functions for diamond and amorphous carbon used in
these calculations are plotted in Figs.\ \ref{f4} and \ref{f5}. In order to 
obtain the ELF for diamond, we have used optical data for diamond
\cite{l24}, \cite{l25} ($E<35$ eV) and X-ray data for scattering of photons 
on carbon \cite{l26} ($E>49.3$ eV). The ELF in the intermediate region $35$ eV $<E<$ $49.3$
eV was fitted in order to fulfill the oscillator-strength sum rule (\ref{oscs}).
The ELF for amorphous carbon was obtained
from optical data \cite{l28} in the region $E\leq40$ eV and from X-ray data on
atomic carbon \cite{l26} ($E>72.4$ eV). As previously, the ELF in the
intermediate region was fitted in order to fulfill the oscillator-strength 
sum rule (\ref{oscs}).
Both diamond and amorphous carbon show dominant peaks in their ELF,
corresponding to well-defined volume plasmons \cite{l19} as expected for
free-electron-like materials. This means that the Lindhard dielectric function
approximation describes these two solids satisfactorily \cite{l14}.

Figs.\ \ref{f2} and \ref{f3} show the IMFPs of electrons in diamond and 
amorphous carbon, calculated from (\ref{ashleyfin}) and (\ref{tanumaf}). 
The IMFPs increase monotonically with impact energy, however the scaling with 
the density of the medium is not preserved explicitly. For low energies 
($E \approx50$ eV), the IMFP shows a characteristic rapid increase, and for the 
TPP-2 model it becomes undefined if approaching the Fermi energy $E_F$ 
(cf.\ (\ref{res2})).

It should be stressed that the first approximations used here give an
upper limit for the total number of secondary electrons liberated by an Auger
electron. We expect therefore, that in reality, the number of these cascade electrons will
be smaller. The present model treats both allotropes of carbon (diamond, an
insulator and amorphous carbon, a conductor) as free-electron-like materials,
and we model their band structure in a free-electron gas approximation 
\cite{l6}. The Fermi energy for diamond is $E_F=28.7$ eV, and for amorphous 
carbon $E_F=21.1$ eV as obtained from the free-electron gas approximation.
We note that the model will give more accurate results by considering the real
band structure of the solid. The Fermi level lies then in the middle of
the band gap at $T=0$ K for semiconductors/insulators.

Based on these initial results, we have constructed a model, which describes 
the time evolution of the secondary electron cascade in diamond and amorphous 
carbon (cf. \cite{l29}, \cite{l30}).
The algorithm for the MC simulation is available from authors.

The model describes the evolution of the cascade in the approximation of
independent non-interacting electrons, neglecting long range Coulomb 
interactions. The latter assumption holds due to the emission
time scales and electron energy ranges relevant for the simulation.
We assume that on average only {\bf one} elastic or inelastic electron-atom 
scattering takes place in a cluster of size $\lambda_{el(in)}$. An electron 
of energy $E$ (cf. Fig.\ \ref{f6}) enters
the solid and undergoes collisions with the atoms. Depending on the magnitude
of the respective cross-sections, either elastic or inelastic collisions occur
as a stochastic process (probability of collision
$\approx\frac{\sigma_{el(in)}}{\sigma_{el}+\sigma_{in}}$).
In elastic collisions, the primary electron travels through the atomic cluster
of size $\lambda_{el}(E)$ and leaves after time
$\Delta t=\frac{\lambda_{el}(E)}{\sqrt{2E}}$. For an inelastic collision the 
situation gets more complicated. First, as previously, the electron travels 
through the atomic cluster of size $\lambda_{in}(E)$. After time 
$\Delta t=\frac{\lambda_{in}(E)}{\sqrt{2E}}$ it loses part of its energy $\om$, 
and transfers it to an electron of energy $E_0$ in the Fermi band ($E_0<E_F$). 
Energy $E_0$ of the electron in the band is chosen,
according to the Fermi density of levels at $T=0$ K \cite{l6} (with no
thermal excitations
assumed).
If the total energy $E_0+\om> E_F$, the secondary electron gets excited, 
and it is emitted instantaneously when the primary electron leaves the 
cluster. Otherwise,
if $E_0+\om < E_F$, the primary electron interacts inelastically with
electrons in the Fermi band, loosing the part of its energy $\om$, however,
no secondary emission occurs in this case. The process continues until
the energies of all excited electrons, including the primary
one, fall below the Fermi barrier $E_F$.

For simplicity we have assumed here that there are no thermal excitations
in the Fermi band ($T=0$ K), and this gives an upper limit of maximal ionization.
If $T>0$, then additional low-occupied energy levels above the Fermi energy become 
available, so the effective energy barrier becomes higher, and cascading will
liberate fewer electrons from the Fermi band.
%%%%%%%%%%%%%%%%%%%%%%%%%%%%%%%%%%%%%%%%%%%%%%%%%%%%%%%%%%%%%%%%%%%%%%%%%%%%%

%%%%%%%%%%%%%%%%%%%%%%%%%%%%%%%%%%%%%%%%%%%%%%%%%%%%%%%%%%%%%%%%%%%%%%%%%%%%%
\section{Numerical results}

MC simulations showed that the number of cascade electrons converged
after five iterations in both samples. A set of 500 simulations was then
performed for each of the two samples in order to obtain a time-dependent
estimate of the number of ionizations. In these simulations, the energy of the
primary electron was fixed at $E=E_F+250$ eV. Cascading included $1+5$
interactions (the primary impact and 5 cascade steps). Figure \ref{f7} shows 
the results.

For diamond the average number of ionization events after the first
femtosecond was estimated to be $\approx6$ based on Ashley's model 
(\ref{ashleyfin}) and $\approx7$ based on the TPP-2 model (\ref{tanumaf}). 
The number of secondary ionizations increased with time, and it saturated 
within about $40$ fs with a total of $37$ electrons released at the maximal
ionization of $\approx37$ events (Ashley). Saturation was slower with the 
TPP-2 model ($100$ fs), and the total number of cascade electrons (about 18) 
was about the half of those ejected in Ashley's model.
It should be stressed that in the latter case (TPP-2) the average number of 
ionizations grew slowly with time. The same scenario held also for the 
cascades in amorphous carbon.
Both Ashley's and the TPP-2 models predicted $6-8$ ionizations after the 
first femtosecond. Calculations based on Ashley's model give a total number of
around $40$ cascade electrons. These electrons were released within the first
$10$ fs, after which no more ionizations occured.
Calculations based on the TPP-2 model level out at $100$ fs, and the total
number of electrons released in the cascade is only about $23$.

The IMFP at $E=250+E_F$ eV calculated from (\ref{ashleyfin}) (Ashley) 
was larger than the corresponding IMFP from the TPP-2 model (\ref{tanumaf}) 
for both diamond and amorphous carbon.
However, the most probable energy loss at this energy is less than
$60$ eV in $80$~\% of the cases as estimated from the integrated energy
loss probability density. This implies that the subsequent cascade is
dominated by secondary electrons of energy $60$ eV and less, and at this
energy, the IMFP calculated in the TPP-2 model is larger than the IMFP in
Ashley's model.
Therefore, the number of ionizations estimated in Ashley's model is larger
for both diamond and amorphous carbon.

We have also plotted the maximal average ionization as the function of
the electron impact energy (cf. Fig.\ \ref{f8}). The total number of 
ionizations increases linearly with impact energy in the energy range between 
$100+E_F$ eV and $300+E_F$ eV, as expected.

In constructing the model, we laid emphasis on formulating a reliable description
of the Auger electron passage through a solid. Therefore we restricted 
ourselves to an estimation of the upper limit of ionizations caused by a single
Auger electron. This approach allowed us to use first-order approximations 
to model electron-solid interactions.
We performed our calculations in the approximation of non-interacting electrons
in the cascade, neglecting long range Coulomb interactions. Since the maximal 
number of ions in the carbonic medium caused by a single primary Auger electron
is small ($\approx20-40$) in comparison with the total number of atoms in the 
sample ($\approx10^9$ atoms for ($100\times100\times100$) nm$^3$ cube),
the approximation of neutral atoms for which the values for the IMFP, EMFP and
DIMFP were derived is supposed to work well. This approach is expected to be useful
for any secondary electron cascade generated by Auger electrons released in 
photoelectric events. 

Moreover, for microscopic samples one may neglect the ionization rate caused by 
a photoelectron, and then approximate the total ionization rate caused 
by a single photoelectric event by the Auger-electron ionization rate. 
This translates into an ionization rate 
of $\approx 20-40$ secondary electrons emitted within the first $100$ 
femtoseconds after the primary electron emission in diamond and amorphous carbon.

Finally, it should be stressed that we modeled the band structure of diamond
and amorphous carbon, using a free-electron-gas approximation. This assumption 
gives an upper estimation of the ionization rate caused by a single Auger 
electron. 
Moreover, the secondary electron emission was considered in case of $T=0$ K 
(with no thermal excitations in the Fermi band), and this, again, overestimates
the maximal number of ionizations. At $T>0$ K, the effective energy barrier becomes 
higher, and cascading will excite fewer electrons from the Fermi band than in the 
case of $T=0$ K.

If one considers the real band structure of the solid, then the expected total
number of electrons ejected in the cascade decreases further as the Fermi level
lies in the middle of the band gap at $T=0$ K (semiconductors/insulators). 
The effect of including the real band structure on the ionization dynamics 
is shown in Fig.\ \ref{f9} for diamond, for which the band structure is well
established \cite{diamond1,diamond2,diamond3}.
%%%%%%%%%%%%%%%%%%%%%%%%%%%%%%%%%%%%%%%%%%%%%%%%%%%%%%%%%%%%%%%%%%%%%%%%%

%%%%%%%%%%%%%%%%%%%%%%%%%%%%%%%%%%%%%%%%%%%%%%%%%%%%%%%%%%%%%%%%%%%%%%%%%
\section{Conclusions}

%%%%%%%%%%%%%%%%%%%%%%%%%%%%%%%%%%%%%%%%%%%%%%%%%%%%%%%%%%%%%%%%%%%%%%%%%%%
The primary photoelectrons and the Auger electrons may escape from very
small samples, however, in larger samples, these electrons may become
trapped and thermalised. This process leads to additional ionization and
to the deposition of further energy into the sample. Thermalisation
involves inelastic electron-atom interactions, and produces secondary
cascade electrons on a longer time scale. In this paper, we analyzed the
specific contribution of Auger electrons to the overall ionization of a
macroscopic sample. The results describe the evolution of Auger-electron
cascades in two model compounds, diamond and amorphous carbon, and show
that a maximum of $20-40$ secondary cascade electrons may be released by a
single Auger electron within the first $100$ femtoseconds following the
emission of the Auger electron. A quantitative description of the
ionization dynamics of target samples is of crucial importance to
practically all planned experiments at X-ray free-electron lasers, ranging
from imaging to the creation of warm dense matter.
%%%%%%%%%%%%%%%%%%%%%%%%%%%%%%%%%%%%%%%%%%%%%%%%%%%%%%%%%%%%%%%%%%%%%%%%%

\section*{Acknowledgments}

We are grateful to Gyula Faigel, Zoltan Jurek, Michel A. van Hove,
Leszek Motyka, Richard Neutze and Remco Wouts for discussions. 
This research has been supported in part by the Polish Committee
for Scientific Research with grants Nos.\ 2 P03B 04718, 2 P03B 05119, 
the EU-BIOTECH Programme and the Swedish Research Councils. 
A.\ S.\ was supported by a STINT distinguished guest professorship. 
B.\ Z.\ was supported by the Wenner-Gren Foundations.

%\bibliographystyle{unsrt}
%\bibliography{secondcd}

\end{document}